\documentclass[useAMS,usenatbib]{mn2e}
\usepackage{psfig}

\input psfig.sty
\usepackage{amsmath}
\usepackage{amssymb}
\usepackage{upgreek}
\usepackage{longtable}
\usepackage[mathscr]{eucal}
\usepackage{graphicx}

\usepackage{color}

\newcommand{\arcs}{\mbox{\ensuremath{^{\prime\prime}}}}

\title
[Ionized Nebula Surrounding the RSG W26]
{The Ionized Nebula surrounding the Red Supergiant W26 in Westerlund 1}
\author
[Wright et al.]
{Nicholas J. Wright$^1$, Roger Wesson$^2$, Janet E. Drew$^1$, Geert Barentsen$^1$, \newauthor 
Michael J. Barlow$^3$, Jeremy R. Walsh$^4$, Albert Zijlstra$^5$, Jeremy J. Drake$^6$, \newauthor
Jochen Eisl{\"o}ffel$^7$ and Hywel J. Farnhill$^1$\\
$^1$Centre for Astrophysics Research, Science and Technology Research Institute, University of Hertfordshire, Hatfield, AL10 9AB, UK\\
$^2$European Southern Observatory, Alonso de C\'ordova 3107, Casilla 19001, Santiago, Chile\\
$^3$Department of Physics and Astronomy, University College London, Gower Street, London, WC1E 6BT, UK\\
$^4$European Southern Observatory, Karl-Schwarzschild-Strasse 2, D-85748 Garching, Germany\\
$^5$School of Physics and Astronomy, University of Manchester, Sackville Street, PO Box 88, Manchester, M60 1QD, UK\\
$^6$Harvard-Smithsonian Center for Astrophysics, 60 Garden Street, Cambridge, MA 02138, USA\\
$^7$Th{\"u}ringer Landessternwarte, Sternwarte 5, D-07778 Tautenburg, Germany\\
}

\begin{document}
\maketitle

\begin{abstract}
We present H$\alpha$ images of an ionized nebula surrounding the M2-5Ia red supergiant (RSG) W26 in the massive star cluster Westerlund~1. The nebula consists of a circumstellar shell or ring $\sim$0.1~pc in diameter and a triangular nebula $\sim$0.2~pc from the star that in high-resolution Hubble Space Telescope images shows a complex filamentary structure. The excitation mechanism of both regions is unclear since RSGs are too cool to produce ionizing photons and we consider various possibilities. The presence of the nebula, high stellar luminosity and spectral variability suggest that W26 is a highly evolved RSG experiencing extreme levels of mass-loss. As the only known example of an ionized nebula surrounding a RSG W26 deserves further attention to improve our understanding of the final evolutionary stages of massive stars.
\end{abstract}


\section{Introduction}

The study of the late stages of stellar evolution of massive stars is fundamentally important for understanding the chemical evolution of galaxies, the role of feedback in the interstellar medium (ISM), and the progenitors of core-collapse supernovae (SNe). Massive stars enrich the ISM with heavy elements during all stages of their evolution, from fast winds in their early lives through cool dense outflows from red supergiants (RSGs) and finally during their deaths in core-collapse SNe explosions. Mass loss is vital to the study of massive stars, not only influencing the derivation of physical properties, but also dictating the evolutionary path and the type of compact remnant left behind.

Direct studies of ejected material from massive stars act as probes of both the evolutionary and mass-loss history of the star \citep[e.g.,][]{barl94}. Despite this, observational examples of ejected circumstellar material around massive stars are rare, and even rarer amongst each of the different evolutionary stages observed with the most thoroughly studied population being the luminous blue variables. Amongst RSGs circumstellar gas has been directly observed around only a few objects \citep[e.g.,][]{schu06,ohna08,maur11}, mostly from radio or sub-mm emission due to neutral gas, or from scattering of optical resonance lines \citep{smit01b}. Ionized gas has never been observed around RSGs since they do not produce ionizing photons.

In this letter we present images of an ionized nebula surrounding the RSG W26 in the young star cluster Westerlund~1. Westerlund~1 is the most massive starburst cluster known in our Galaxy, with a total stellar mass of $5 \times 10^4$~M$_\odot$ \citep{bran08} and more than 50 known massive O-type stars, Wolf-Rayet stars and RSGs at various stages of post-main sequence evolution \citep{clar05b}. Despite being discovered more than 50 years ago \citep{west61} it has been poorly studied since then due to the high interstellar reddening in its direction \citep[$A_V \sim 11$--13~mag,][]{clar05b}, though it lies at a distance of only $3.55 \pm 0.17$~kpc \citep{bran08}. Spectra of W26 suggest a spectral type of M2-5Ia \citep{clar10} and have continually exhibited nebular emission lines \citep{west87,clar05b}. Images of W26 have revealed extended mid-IR \citep{clar98} and radio emission \citep{clar98,doug10} in its vicinity.

\begin{figure}
\begin{center}
\includegraphics[width=7.8cm]{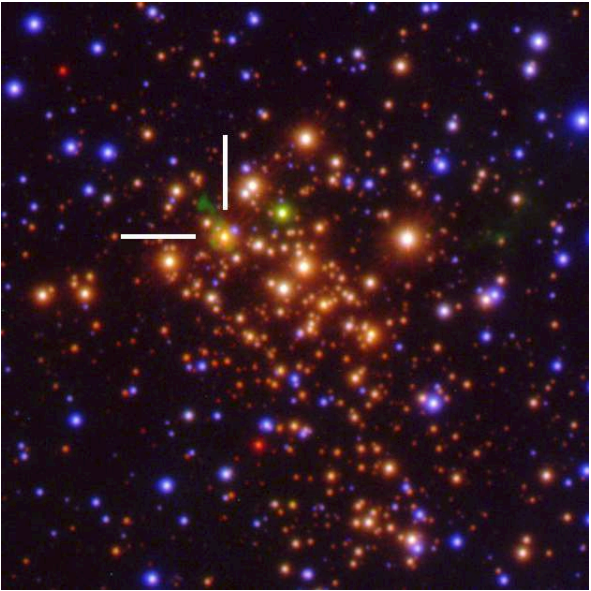}
\caption{VPHAS+ 3-colour image of Westerlund~1 composed of data from the $g$ (blue), H$\alpha$ (green), and $i$ (red) filters. The image is $2.5^\prime \times 2.7^\prime$, equivalent to $2.6 \times 2.8$~pc at a distance of 3.55~kpc. Ionized nebulae stand out clearly in green in this image with the nebula surrounding W26 indicated by the white lines.}
\label{3colour}
\end{center}
\end{figure}

In this paper we present definitive H$\alpha$ images that confirm an ionized nebula surrounding W26. The images were obtained as part of the VLT (Very Large Telescope) Survey Telescope (VST) Photometric H$\alpha$ Survey (VPHAS+, Drew et al. 2013, {\it in prep.}) and reveal a detached circumstellar nebula surrounding W26 with a triangular ionized gas feature 10\arcs\ from the star. In Section~2 we present the VPHAS+ observations as well as archival data. In Section~3 we discuss the nebulous structures and make comparisons with other evolved massive stars with circumstellar nebulae.

\section{Observations}

\subsection{VPHAS+ H$\alpha$ Observations}

VPHAS+ is a survey of the southern Galactic plane in broad-band Sloan $u$, $g$, $r$, $i$, and narrow-band H$\alpha$ filters using OmegaCam on the 2.6m VST. Tiling of the 1~deg$^2$ OmegaCam field-of-view leads to almost uninterrupted coverage across the Galactic plane at high spatial resolution (1 pixel = 0.21\arcs). The use of a narrow-band H$\alpha$ filter ($\lambda_c = 6588$ \AA, FWHM = 107 \AA) picks out extended emission-line nebulae and point sources as the northern hemisphere counterpart survey IPHAS \citep{drew05} also did \citep[e.g.,][]{with08,wrig12,sabi13}.

\begin{figure*}
\begin{center}
\includegraphics[width=15.0cm]{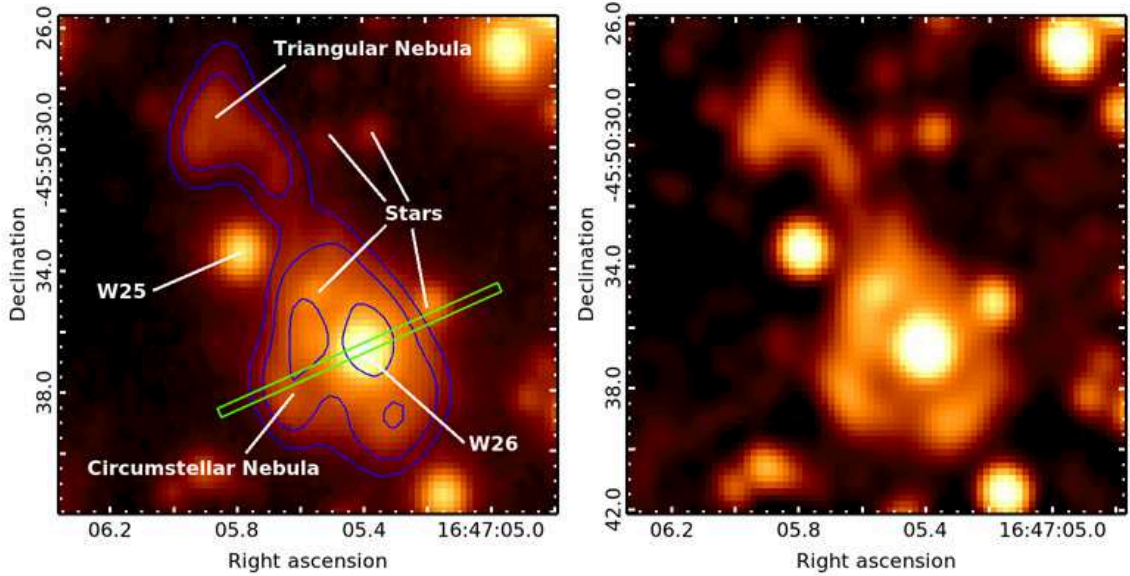}
\caption{VPHAS+ H$\alpha$ images of the nebula surrounding W26, scaled logarithmically. {\it Left:} Original image with labels indicating the nebular components, stars identified in the VPHAS+ $i$-band image, W25 and W26. Overlaid in blue are 8.6~GHz contours from \citet{doug10} and in green the VLT/FORS slit shown in Figure~\ref{fors_profile}. {\it Right}: Deconvolved image showing the detached circumstellar nebula.}
\label{vphas_greyscale}
\end{center}
\end{figure*}

The field including Westerlund~1 was observed as part of VPHAS+ on 30 April and 11 May 2012 and a 3-colour image of the cluster is shown in Figure~\ref{3colour}. The seeing was 0.8\arcs\ as measured on the H$\alpha$ images. Excess H$\alpha$ emission stands out in green on this image. There is evidence for excess H$\alpha$ emission in other sources in this image, but it does not appear extended as it does around W26.

Figure~\ref{vphas_greyscale} shows an H$\alpha$ image centered on W26. Two distinct morphological components are clear: circumstellar emission surrounding the star and a triangular nebula $\sim$10\arcs\ to the north-east. Figure~\ref{vphas_greyscale} also shows a deconvolved H$\alpha$ image created using a Lucy-Richardson \citep{lucy74} deconvolution algorithm and employing a point spread function (PSF) created from 15 isolated field stars. In this image the circumstellar nebula appears detached from W26 allowing the possibility that the nebula is an edge-brightened circumstellar shell or ring surrounding the star. The structure appears fragmented with multiple bright spots at distances of 2.4--3.2\arcs\ from the star (0.04--0.06~pc at $d = 3.55$~kpc). Some of the bright spots are identified as stars from the VPHAS+ $i$-band image, although the majority are nebulous.

\subsection{Hubble Space Telescope Observations}

\begin{figure*}
\begin{center}
\includegraphics[height=6.3cm]{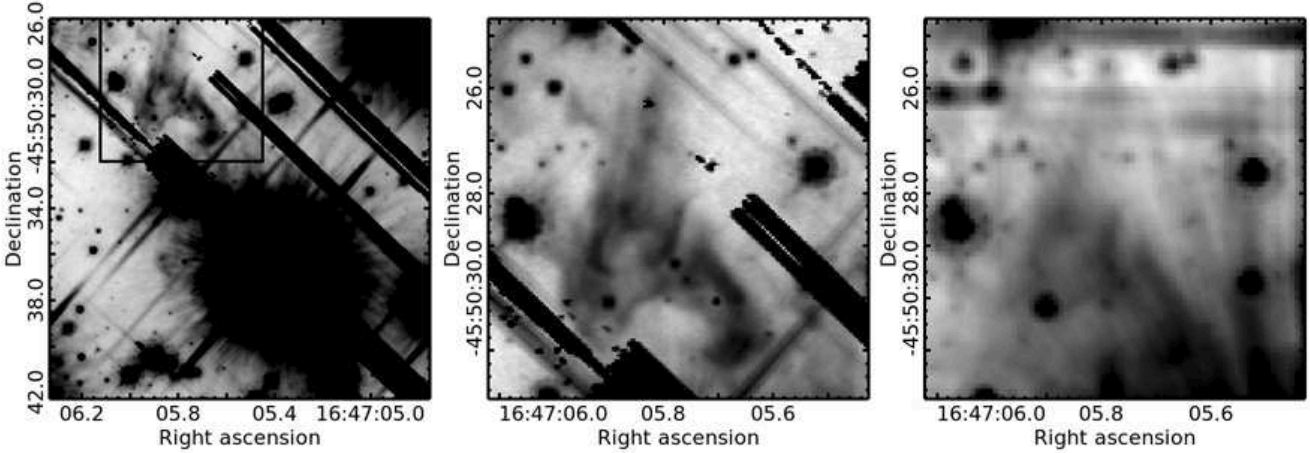}
\caption{Greyscale images of the triangular nebula seen with HST. {\it Left:} WFC/ACS image with the F814W filter (a broad $I$-band filter). The black box indicates the area shown in the other two panels. {\it Centre:} ACS/F814W filter image centred on the triangular nebula. {\it Right:} WFC3/F125W filter image (a broad $J$-band filter).  All the images are inverted and logarithmically scaled.}
\label{three_panels}
\end{center}
\end{figure*}

Westerlund~1 was observed by the Hubble Space Telescope (HST) on 2005 June 26 (programme no. 10172) with the Advanced Camera for Surveys (ACS, 0.05\arcs/pixel) Wide Field Channel (WFC) using the F814W filter ($\lambda_c = 8333$ \AA, $\Delta \lambda = 2511$ \AA, a broad $I$ filter). Three exposures were made for a total of 2407s, shown in Figure~\ref{three_panels}. The long exposure time means the vast majority of stars are saturated, with diffraction spikes dominating the centre of the cluster. The nebula surrounding W26 is not detected due to the extended wings of the PSF from W26, but the triangular nebula is clearly resolved and shows a complex filamentary structure.

The cluster was observed again on 2010 August 25--27 (programme no. 11708) using the Wide Field Camera 3 (WFC3) IR channel (0.13\arcs/pixel) using the broad-band near-IR filters F125W, F139M, and F160W. As with the ACS images the circumstellar nebula around W26 is obscured, while the triangular nebula is only barely visible in the F125W filter images ($\lambda_c = 1.25 \mu$m, $\Delta \lambda = 0.28 \mu$m, approximately a $J$-band filter) and obscured by W26 in the longer wavelength images. Seven exposures were made, for a total of 2445s, also shown in Figure~\ref{three_panels}.

\subsection{VLT/FORS Spectroscopy}
\label{s-spectroscopy}

W26 was observed with the FOcal Reducer and low dispersion Spectrograph (FORS) on the VLT on 2011 April 16 (observation ID 539166) with a slit rotated to a position angle of -113$^\circ$ covering the star and the circumstellar nebula (Fig~\ref{vphas_greyscale}) and covering 5800--7200\AA. The spectrum of W26 is consistent with previous estimates of the spectral type and does not reveal any emission lines from the star itself. The nebular emission includes lines of [N~{\sc ii}], H$\alpha$, [S~{\sc ii}], He~{\sc i}, [Ar~{\sc iii}] and [S~{\sc iii}], though many of these are not measurable with sufficient accuracy to provide density or temperature diagnostics for the gas. The spectral resolution is also not high enough to measure velocity differences between the star and the nebula of the order of a few km/s. The current observations can only set an upper limit for the nebular expansion of $\sim$30~km/s, which is consistent with RSG mass-loss.

\begin{figure}
\begin{center}
\includegraphics[height=8cm,angle=270]{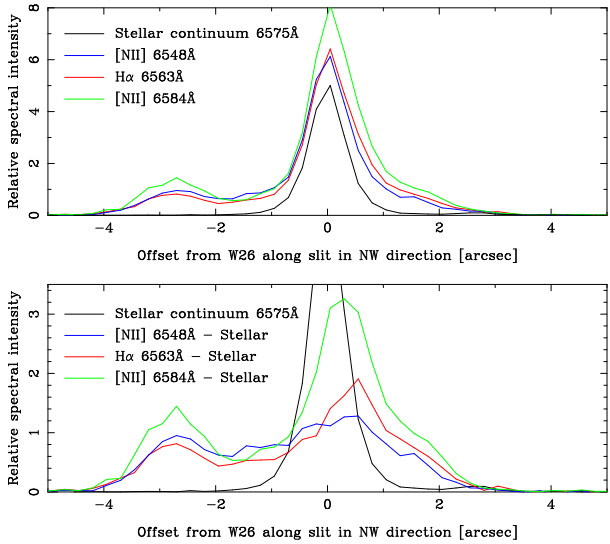}
\caption{Spatially resolved emission along the VLT/FORS slit (as shown in Fig~\ref{vphas_greyscale}) at the [N~{\sc ii}] 6548, 6584 and H$\alpha$ emission lines, as well as at 6575\AA\ to show the stellar continuum emission. The upper panel shows absolute measurements while the lower panel shows the emission lines with the stellar continuum subtracted (as well as the stellar continuum profile itself for reference).}
\label{fors_profile}
\end{center}
\end{figure}

The [N~{\sc ii}] (6548+6584) / H$\alpha$ ratio of 2.94 is very high and likely implies CNO-cycle enrichment, as has been observed in the nebulae surrounding other evolved massive stars \citep{john92}. The emission lines seen in the FORS spectrum confirm that the circumstellar nebula is ionized and that the emission is not due to reflection or continuum emission. Figure~\ref{fors_profile} shows the emission along the slit in the H$\alpha$ and [N~{\sc ii}] emission lines, confirming that the nebula is extended with evidence of edge-brightening (the nebular emission peaks $\sim$3\arcs\ south-east and $\sim$0.5\arcs\ north-west of W26). The H$\alpha$ and [N~{\sc ii}] 6584\AA\ lines likely dominate extended emission in the VPHAS+ H$\alpha$ filter. Emission in the HST F814W filter is probably a combination of [O~{\sc ii}] 7320, 7330\AA, [S~{\sc iii}] 9069, 9533\AA\ \citep[as seen in the FORS spectrum and the stellar spectrum of W26 by][]{clar05b} and Paschen series lines, while emission in the HST F125W filter is likely to be dominated by Pa$\beta$ \citep[see][]{weid13}.

\section{Discussion}

\subsection{The red supergiant W26}

W26 is known to be spectrally variable with changes in the strength of the TiO bandheads on timescales of years resulting in a spectral type range of M2-5Ia \citep{clar10}, making it one of the latest RSGs known \citep{leve05}. Spectral variability is not a common feature amongst RSGs that \citet{mass07} suggest may be caused by the star being out of hydrostatic equilibrium in the Hayashi ``forbidden zone'' that also results in extensive mass-loss. From the observed $K$-band magnitude of 1.9 \citep{meng07}, extinction of $A_K = 1.13$ \citep{bran08} and a bolometric correction of $BC_K = 2.79$ (2.96) mag for type M2I (M5I) \citep{leve05}, W26 has $M_{bol} = -9.19$ (-9.02), $L_{bol} = 3.8 \times 10^5$ ($3.2 \times 10^5$) L$_\odot$. This is remarkably similar to the M5 supergiant VY~CMa, one of the most luminous known RSGs with $L = 3.5 \times 10^5$~L$_\odot$ \citep[at $d = 1.14$~kpc,][]{choi08}. Comparing W26 to the \citet{leve05} sample of Galactic RSGs with luminosities determined from the $K$-band (their Fig~3$c$) it is clear that it is one of the latest and most luminous RSGs known in our Galaxy. It is also one of the largest, with a radius of 1530 (1580) $R_\odot$, similar to the massive RSG, WOH~G64, which notably is the only other RSG observed to have an emission line spectrum \citep{leve09}. This is all consistent with W26 being a very evolved RSG with a greater level of instability and higher mass-loss rate than typical RSGs.

Ionized nebulae have never previously been resolved around RSGs as their photospheres are too cool to produce ionizing photons and other ionizing sources must be sought. The strong [S~{\sc iii}] emission lines observed by \citet{clar05b} suggest either a very high temperature H~{\sc ii} region or shock ionization of the circumstellar nebula. If the nebula is photoionized it could be due to either the radiation field of the cluster or a hot companion. There is no evidence for the latter however, and the small radial velocity variations observed by \citet{cott12} are consistent with pulsations. The current RSG wind is at least partly molecular as SiO and H$_2$O masers are observed towards W26 with velocities consistent with a RSG wind \citep{fok12}. The authors also note that W26 is a significantly less intense maser source than other nearby and isolated RSGs (such as VY~CMa), which they argue could be due to the outer layer of the molecular envelope being disturbed by the cluster environment in Westerlund~1.

\subsection{The circumstellar nebula}

The circumstellar nebula, only clearly resolved in the VPHAS+ observations, appears to be fully detached from W26 in the deconvolved image. It is difficult to determine the morphology of the nebula given the limitations of the available observations, which suggest a clumpy and inhomogeneous structure, possibly indicating some degree of fragmentation. The nebula could be either an edge-brightened circumstellar shell or a ring, and is reminiscent of the complex, if more compact, circumstellar reflection nebula surrounding VY~CMa ($\sim$0.04~pc in diameter compared to $\sim$0.1~pc for W26's nebula). \citet{smit01b} estimate the mass of the nebula surrounding VY~CMa as 0.2--0.4~M$_\odot$, very similar to the ionized mass of 0.26~M$_\odot$ estimated by \citet{doug10} for the nebula surrounding W26.

We performed aperture photometry on the VPHAS+ H$\alpha$ image and estimate a total flux of the circumstellar nebula of $2.8 \times 10^{-13}$~ergs/cm$^2$/s over an area of 31 arcsec$^2$, a surface brightness of $\sim$$10^{-14}$~ergs/cm$^2$/s/arcsec$^2$. Since the VPHAS+ H$\alpha$ filter includes the [N~{\sc ii}] 6548\AA\ and 6584\AA\ lines (the former at $\sim$75\% transmission) we use the relative strengths of these lines in the VLT/FORS spectrum to calculate an H$\alpha$ flux of $7.5 \times 10^{-14}$~ergs/cm$^2$/s. This is probably overestimated due to contamination from the PSF wings of W26 and nearby stars superimposed on the nebula.

\subsection{The triangular nebula}

The triangular nebula, $\sim$0.2~pc from W26, is most clearly resolved in the HST images and shows a filamentary structure with a size of $\sim$4.5\arcs\ or 0.08~pc. The filaments in the triangular nebula appear to be pointed towards the nearby blue supergiant (BSG) W25 \citep{negu10}, possibly suggesting the triangular nebula is associated with this star instead of W26. However the deconvolved VPHAS+ images hints at a thin nebulous strip that might be connecting the triangular nebula with the circumstellar nebula around W26 \citep[also seen in the 8.6-GHz radio images,][]{doug10}. This connection could suggest that the nebula is associated with W26, although it may be shaped or photoionized by W25 if they are at similar distances within Westerlund~1.

The triangular nebula could be some sort of outflow from W26, possibly along a polar axis and similar to the circumstellar ring and bipolar outflows around the BSG Sher~25. The ring nebula around Sher~25 has a diameter of 0.4~pc and an outflow 0.5~pc in length \citep{bran97a}, slightly larger than the structures seen around W26. There is some uncertainty as to whether the nebula around Sher~25 was ejected during the star's current BSG phase or a previous RSG phase \citep[e.g.,][]{smar02} and the detection of similar structures around a RSG might help resolve this. While Sher~25 does have two polar outflows, one side is notably brighter than the other, which might suggest that a second outflow exists around W26 but is not detected. Alternatively outflows may exist on both sides of W26 but are only being photoionized on one side by the BSG W26. The possibility that the non-symmetry is due to variable extinction across the cluster can be ruled out by the fact that a corresponding structure is not seen in the radio images.

Alternatively the triangular nebula could be a flow of material being channeled off the circumstellar nebula by the cumulative cluster wind and radiation field. \citet{muno06} find evidence for a hot intra-cluster medium in Westerlund~1 that could cause this and would also promote charge-exchange ionization with cool gas ejected from W26. This theory is supported by the lack of nebulous emission on the opposite side of W26 and the alignment of the structure pointing away from the centre of Westerlund~1 \citep[see discussion in][]{doug10}, though the narrowness of the strip possibly connecting the triangular nebula with the circumstellar nebula suggests some degree of collimation more consistent with a polar outflow than with a stripping effect.

Aperture photometry of the triangular nebula on the VPHAS+ image gives a total flux of $7.2 \times 10^{-14}$~ergs/cm$^2$/s over an area of 23 arcsec$^2$, a surface brightness of $\sim$$3 \times 10^{-15}$~ergs/cm$^2$/s/arcsec$^2$. Accounting for the [N~{\sc ii}] lines this gives an H$\alpha$ flux of $7.2 \times 10^{-14}$~ergs/cm$^2$/s. This can be compared with the 4.8~GHz radio flux from the triangular nebula of 41.5~mJy \citep{doug10}, which has a flat radio spectrum consistent with optically thin free-free emission from an ionized nebula. Using eqn.~6 of \citet{miln75} and assuming a temperature of 8,000~K and a ratio of He$^+$/H$^+ = 0.1$ this implies an unreddened H$\beta$ flux of $1.5 \times 10^{-11}$~ergs/cm$^2$/s\footnote{This equation is mildly sensitive to these parameters. A 10\% increase in the nebular temperature produces a $\sim$6\% decrease in the resulting H$\beta$ flux, while increasing the He$^+$/H$^+$ ratio to 0.15 results in the H$\beta$ flux decreasing by $\sim$4\%.}, which for Case~B recombination gives an unreddened H$\alpha$ flux of $4.4 \times 10^{-11}$~ergs/cm$^2$/s. Comparing this to the observed H$\alpha$ flux gives an extinction of $A_{H\alpha} = 8.4$~mags. Assuming $R_V = 3.1$ \citep{howa83} this implies an extinction of $A_V = 10.5$~mags, lower than the typical range of $A_V = 11.6$--13.6~mag found by \citet{clar05b} for Westerlund~1. This is reconcilable if the extinction law is non-standard, e.g., \citet{bran08} suggest $R_V = 3.7$ for which we calculate $A_V = 12.5$~mags, in good agreement with existing measurements.

\section{Summary}

We have presented VPHAS+ H$\alpha$ observations that establish the existence of an ionized circumstellar nebula surrounding the RSG W26 in Westerlund~1. The nebula consist of a detached circumstellar shell or ring surrounding the star and a triangular nebula located $\sim$0.2~pc from the star, possibly connected to the ring via a thin nebulous strip or outflow. The excitation mechanism of both regions is unclear, since RSGs are too cool to produce ionizing photons. They may be being photoionized by either a hot companion to W26, the nearby BSG W25, the cluster radiation field, or even shock excited due to collisions with the intra-cluster medium.

The presence of the nebula suggests extensive mass-loss in the recent history of W26. Its late spectral type, very high luminosity and spectral variability all suggest the star to be highly evolved amongst the RSGs. Both the star and the nebula are comparable to the RSGs VY~CMa and WOH~G64, both of which are highly luminous late-type RSGs with evidence for circumstellar gas.

W26 provides a rare opportunity to directly investigate an extreme mass-loss event from a highly evolved RSG. To follow-up on this discovery high resolution narrow-band imaging could be used to resolve the nebular structures better or spectroscopy of the ionized nebula could provide insights into the physical and dynamical conditions in the gas and help resolve the origin of both structures. Higher resolution spectra are also necessary to measure velocity differences between the star and the nebula to determine the expansion velocity of the nebula.

\vspace{-0.5cm}

\section{Acknowledgments}

We are grateful to the anonymous referee for a helpful report that has improved the content of this paper. These observations were obtained as part of VPHAS+, an ESO Public Survey, and processed by the Cambridge Astronomical Survey Unit. We acknowledge data products from the NASA/ESA HST. We are grateful to Sean Dougherty for supplying the 8.6~GHz radio data shown in Figure~2 and to Chris Evans for comments and discussions on this work. NJW acknowledges a Royal Astronomical Society Research Fellowship. RW acknowledges funding from the Marie Curie Actions of the European Commission (FP7-COFUND).

\vspace{-0.7cm}

\bibliographystyle{mn2e}
\bibliography{/Users/nwright/Documents/Work/tex_papers/bibliography.bib}
\bsp

\end{document}